\documentclass[journal=jctcce,manuscript=article]{achemso}
\setkeys{acs}{usetitle=true}

\usepackage[version=3]{mhchem}
\usepackage{graphicx,color,subfigure}
\usepackage{epstopdf}  
\usepackage{grffile}   
\usepackage{comment}
\usepackage{placeins}
\usepackage{siunitx}
\usepackage{sectsty}
\usepackage{balance}
\usepackage{xspace}
\usepackage{algorithm,algorithmic}
\usepackage{epsfig,amssymb,amsmath,amsthm,amsfonts,amsbsy,mathrsfs,bm}
\usepackage{epstopdf}
\usepackage{threeparttable}
\usepackage{lastpage}
\usepackage{multicol}
\usepackage{multirow}
\usepackage[format=plain,justification=raggedright,singlelinecheck=false,font=small,labelfont=bf,labelsep=space]{caption}

\allowdisplaybreaks
\usepackage{lineno}
\usepackage[version=3]{mhchem}

\title[Gradient]
{V2Rho-FNO: Fourier Neural Operator for Electronic Density Prediction}

\author{Yingdi Jin\textsuperscript{\#}}
\email{jinyd@ustc.edu.cn} 
\affiliation{State Key Laboratory of Precision and Intelligent Chemistry, University of Science and Technology of China, Hefei, Anhui 230026, China}

\author{Xinming Qin\textsuperscript{\#}}
\affiliation{Hefei National Research Center for Physical Sciences at the Microscale, University of Science and Technology of China, Hefei, Anhui 230026, China} 

\author{Ruichen Liu}
\affiliation{Hefei National Research Center for Physical Sciences at the Microscale, University of Science and Technology of China, Hefei, Anhui 230026, China} 

\author{Jie Liu}
\email{liujie86@ustc.edu.cn} 
\affiliation{Hefei National Laboratory, University of Science and Technology of China, Hefei 230088, China}

\author{Zhenyu Li}
\email{zyli@ustc.edu.cn}
\affiliation{Hefei National Laboratory, University of Science and Technology of China, Hefei 230088, China}
\alsoaffiliation{State Key Laboratory of Precision and Intelligent Chemistry, University of Science and Technology of China, Hefei, Anhui 230026, China} 

\author{Jinlong Yang}
\email{jlyang@ustc.edu.cn}
\affiliation{Hefei National Laboratory, University of Science and Technology of China, Hefei 230088, China}
\alsoaffiliation{State Key Laboratory of Precision and Intelligent Chemistry, University of Science and Technology of China, Hefei, Anhui 230026, China}

\begin{document}
\begin{abstract}
Density functional theory (DFT) is a cornerstone of modern computational chemistry and materials science; nevertheless, its computational expense limits its deployment in large-scale and high-throughput applications that rely on explicit electronic-structure calculations. While machine learning methods have accelerated energy calculations for specific molecular categories\cite{Butler2018,Schutt2018_SchNet,Unke2021_SpookyNet}, achieving transferable electronic density predictions across diverse chemical spaces remains a significant challenge\cite{Brockherde2017,Jorgensen2022_eGNN,Fiedler2023_LengthScale}. This study introduces a universal framework based on Fourier Neural Operators (FNOs), which directly learns mappings from external potentials to electronic density distributions\cite{Li2021_FNO,Kovachki2023_JMLRNeuralOperator}. After training on datasets containing various elements and molecular configurations, our model achieves zero-shot generalization to completely unknown molecular systems, accurately predicting their electronic densities. This transferability stems from FNOs' inherent capability to capture global electronic interactions without explicit atomic orbitals or basis sets\cite{Li2021_FNO,Kovachki2023_JMLRNeuralOperator}. The proposed method opens new avenues for rapid and accurate electronic structure calculations, providing a powerful tool for exploring chemical space. 
Density functional theory (DFT) is a cornerstone of modern computational chemistry and materials science; nevertheless, its computational expense limits its deployment in large-scale and high-throughput applications that rely on explicit electronic-structure calculations. While machine learning methods have accelerated energy calculations for specific molecular categories\cite{Butler2018,Schutt2018_SchNet,Unke2021_SpookyNet}, achieving transferable electronic density predictions across diverse chemical spaces remains a significant challenge\cite{Brockherde2017,Jorgensen2022_eGNN,Fiedler2023_LengthScale}. This study introduces a universal framework based on Fourier Neural Operators (FNOs), which directly learns mappings from external potentials to electronic density distributions\cite{Li2021_FNO,Kovachki2023_JMLRNeuralOperator}. After training on datasets containing various elements and molecular configurations, our model achieves zero-shot generalization to completely unknown molecular systems, accurately predicting their electronic densities. This transferability stems from FNOs' inherent capability to capture global electronic interactions without explicit atomic orbitals or basis sets\cite{Li2021_FNO,Kovachki2023_JMLRNeuralOperator}. The proposed method opens new avenues for rapid and accurate electronic structure calculations, providing a powerful tool for exploring chemical space. 
\end{abstract}

\section{Introduction}

Density functional theory (DFT) has become the cornerstone of modern electronic structure calculations due to its favorable balance between accuracy and computational efficiency \cite{Hohenberg1964,Kohn1965}. Central to DFT is the Hohenberg--Kohn (HK) theorem, which establishes that the ground-state electron density uniquely determines all properties of an interacting many-electron system and, in particular, is uniquely determined by the external potential up to an additive constant \cite{Hohenberg1964}. In practical Kohn--Sham formulations, the electron density is obtained through a self-consistent solution of effective single-particle equations \cite{Kohn1965}. Despite its success, the associated self-consistent field (SCF) procedure remains computationally demanding for large systems, high-throughput screening, and time-dependent simulations \cite{Payne1992_RMP,Vanderbilt1990_USPP}.

Recent years have witnessed rapid progress in machine learning approaches aimed at accelerating or partially bypassing electronic structure calculations \cite{Butler2018,Schutt2018_SchNet}. A particularly active direction focuses on learning electron densities directly, either to replace SCF iterations or to enable fast downstream property evaluation \cite{Snyder2012_PRL,Brockherde2017,Jorgensen2022_eGNN,Fiedler2023_LengthScale}. These approaches span kernel methods, deep neural networks, and equivariant graph neural networks, achieving impressive accuracy across molecules, liquids, and solids 
\cite{Unke2021_SpookyNet,Jorgensen2022_eGNN}.
%\cite{Unke2021_SpookyNet,Bhowmik2022,Jorgensen2022_eGNN}.
In most existing formulations, however, density prediction is posed as a mapping from discrete atomic descriptors---such as element types and nuclear coordinates---to continuous density fields. As a consequence, substantial architectural effort is required to encode physical symmetries, long-range interactions, and resolution dependence through handcrafted embeddings, message-passing schemes, or equivariant constraints.

A conceptually more fundamental alternative is to learn the HK map itself, namely the functional mapping from the external potential to the ground-state electron density. Several studies have explored this idea in reduced settings, including one-dimensional model systems and molecular datasets, demonstrating that the HK map can be approximated by machine learning models with encouraging accuracy \cite{Brockherde2017,Bai2022_HKExcited}. Nevertheless, extending this approach to fully three-dimensional systems remains challenging, particularly when accurate densities must be resolved on fine spatial grids. Moreover, most existing models are intrinsically tied to a fixed spatial discretization, limiting their ability to generalize across resolutions and system representations.

In parallel, the development of neural operators has introduced a new paradigm for learning mappings between function spaces \cite{Kovachki2023_JMLRNeuralOperator}. Unlike conventional neural networks, which learn finite-dimensional input--output relations, neural operators are designed to approximate operators acting between infinite-dimensional function spaces. A key consequence of this formulation is that, once trained, a neural operator can be evaluated on inputs represented on discretizations different from those used during training, provided that the underlying functions lie within the operator’s domain of approximation \cite{Kovachki2023_JMLRNeuralOperator,DeHoop2022_OperatorLearning}. Among these methods, the Fourier neural operator (FNO) has proven particularly effective for learning nonlocal operators by combining global spectral convolutions with local nonlinear transformations \cite{Li2021_FNO}. While neural operators have been widely applied to explicit partial differential equations, their potential for learning implicit physical maps in quantum many-body problems has only recently begun to be explored \cite{Kim2024_GPWNO}.

In this work, we propose \emph{V2Rho-FNO}, a three-dimensional Fourier neural operator that directly learns the external-potential-to-density mapping for electronic systems. Unlike structure-based density models that rely on discrete atomic descriptors, our approach takes the nuclear electrostatic potential as the sole input and predicts the ground-state electron density on uniform spatial grids. This formulation is directly motivated by the Hohenberg--Kohn theorem \cite{Hohenberg1964} and treats both input and output as continuous physical fields, placing the problem naturally within the neural operator framework \cite{Kovachki2023_JMLRNeuralOperator,Li2021_FNO}.

From the perspective of operator learning, V2Rho-FNO aims to approximate a \emph{universal} functional relationship between external potentials and ground-state densities, rather than a mapping tied to specific molecular structures. In neural operator theory, generalization is governed not by similarity between individual training and test samples, but by coverage of the input function space on which the operator is defined \cite{Kovachki2023_JMLRNeuralOperator,Lanthaler2022_ConvergenceNO}. Consequently, once sufficient and diverse training data are provided to span the relevant space of external potentials, a trained operator is expected to generalize to previously unseen molecular systems without requiring structural similarity to the training set. This property aligns naturally with the universality implied by density functional theory and distinguishes operator-based learning from structure-centric machine learning models.

An important practical consequence of the operator formulation is discretization invariance. Because Fourier neural operators learn a truncated spectral representation of a continuous operator, the learned mapping can be evaluated on grids of different resolutions without retraining \cite{Li2021_FNO,Kovachki2023_JMLRNeuralOperator}. In practice, resolution transfer is achieved by evaluating the learned operator in Fourier space and extending the spectral representation to finer grids through zero-padding before applying the inverse Fourier transform. This procedure corresponds to a band-limited extension of the learned continuous operator, yielding physically meaningful low-frequency density predictions on finer grids. As such, super-resolved densities should be interpreted as consistent continuations of the learned operator, rather than exact reconstructions of all high-frequency features, in line with recent studies on density super-resolution \cite{Li2025_DensitySR}.

A key conceptual distinction between the present work and most existing 
machine-learning approaches for electronic structure lies in the nature of 
generalization. The majority of first-principles-inspired machine learning models, 
including kernel methods, deep neural networks, and equivariant graph-based models, 
are trained to approximate target quantities as functions of discrete atomic 
descriptors such as nuclear species and coordinates %\cite{Snyder2012_PRL,Unke2021_SpookyNet,Jorgensen2022_eGNN}. 
%\cite{Snyder2012_PRL,Unke2021_SpookyNet,Bhowmik2022,Jorgensen2022_eGNN}. 
In these formulations, generalization is inherently empirical: accurate predictions 
for unseen systems rely on the similarity of local chemical environments, bonding 
patterns, or structural motifs between the training and test data. Consequently, 
transferability is typically achieved through interpolation within a carefully 
curated training distribution, and predictive performance degrades as the target 
systems depart from the chemical or configurational space represented in the data.

Neural operators adopt a fundamentally different learning objective. Rather than 
approximating a finite-dimensional input--output function, neural operators aim to 
learn an operator acting between infinite-dimensional function spaces 
\cite{Kovachki2023_JMLRNeuralOperator}. In this setting, the training data consist of 
samples of input and output functions, and the learned model represents an 
approximation to an underlying continuous operator. As a result, generalization is 
not governed by the proximity of individual samples in descriptor space, but by the 
extent to which the training set provides sufficient coverage of the relevant input 
function space on which the operator is defined 
\cite{Lanthaler2022_ConvergenceNO,DeHoop2022_OperatorLearning}. 

This distinction is particularly important for electronic structure problems. When 
learning the Hohenberg--Kohn map, the target object is not a system-specific mapping 
tied to particular molecular identities, but a universal functional that assigns a 
ground-state electron density to any physically admissible external potential 
\cite{Hohenberg1964}. From the operator-learning perspective, once a neural operator 
has been trained on a sufficiently rich ensemble of external potentials spanning the 
relevant function space, it is expected to generalize to previously unseen molecular 
systems without requiring structural similarity to the training set. In this sense, 
the zero-shot transferability of V2Rho-FNO arises from the operator-level formulation 
itself, rather than from empirical pattern matching across molecular structures.

By learning a mapping between fields rather than between discrete structures and densities, V2Rho-FNO admits a clear physical interpretation: the learned neural operator approximates a generalized, nonlinear density response to the external potential. Together, these features establish V2Rho-FNO as a physically grounded and scalable route toward fast electron density prediction, consistent with the foundational principles of density functional theory \cite{Hohenberg1964,Kohn1965}.

A conceptually useful way to formalize the Hohenberg--Kohn (HK) theorem is as an operator-level map between function spaces,
$\mathcal{G}: V_{\mathrm{ext}}(\mathbf r)\mapsto \rho(\mathbf r)$.
In this perspective, the HK map is intrinsically a \emph{response-type} object: wherever the map admits a (Fr\'echet/G\^ateaux) derivative, its local linearization around a reference potential $V_0$ takes the familiar linear-response form
$\delta\rho(\mathbf r)=\int \chi_{V_0}(\mathbf r,\mathbf r')\,\delta V(\mathbf r')\,d\mathbf r'$,
i.e., the derivative of $\mathcal{G}$ is precisely the static density response operator $\chi$ \cite{Giuliani2005_ElectronLiquid,Ullrich_TDDFT,MarquesGross2004_TDDFT}.
Recent foundational work has clarified the structure (and subtleties) of the density--potential mapping and its differentiability in exact DFT, and has provided differentiable-but-exact regularized formulations where such derivatives are well-defined \cite{Penz2022_StructureHK,Kvaal2013_DifferentiableDFT}.
This makes neural operators particularly well matched to the HK learning objective: neural operators are designed to approximate nonlinear operators between function spaces in a discretization-invariant manner, with generalization governed by coverage of the input function space rather than similarity between individual samples \cite{Kovachki2023_JMLRNeuralOperator}.

\section{Results}

\subsection{Generalization under different training--test relations}

We first examine the predictive behavior of V2Rho-FNO under three increasingly
challenging training--test relations, designed to probe different levels of
generalization difficulty.

\paragraph{(a) Interpolation along molecular dynamics trajectories.}
For single-molecule molecular dynamics trajectories, a contiguous subset of
configurations is used for training, while the remaining configurations are
reserved for testing. This setting corresponds to interpolation along a
low-dimensional configurational manifold defined by nuclear motion at fixed
chemical composition. For both the water and benzene trajectories, V2Rho-FNO
achieves extremely small test losses and near-perfect density correlations with
reference DFT results. The predicted densities accurately reproduce subtle
variations along the trajectories, indicating that the learned operator can
faithfully capture smooth deformations of the external potential induced by
nuclear motion. As expected, this scenario represents the easiest generalization
regime, where training and test potentials densely sample a narrow region of the
input function space.

\begin{figure}[!htb]   	
\centering	
\includegraphics[width=1.0\linewidth, scale=1]{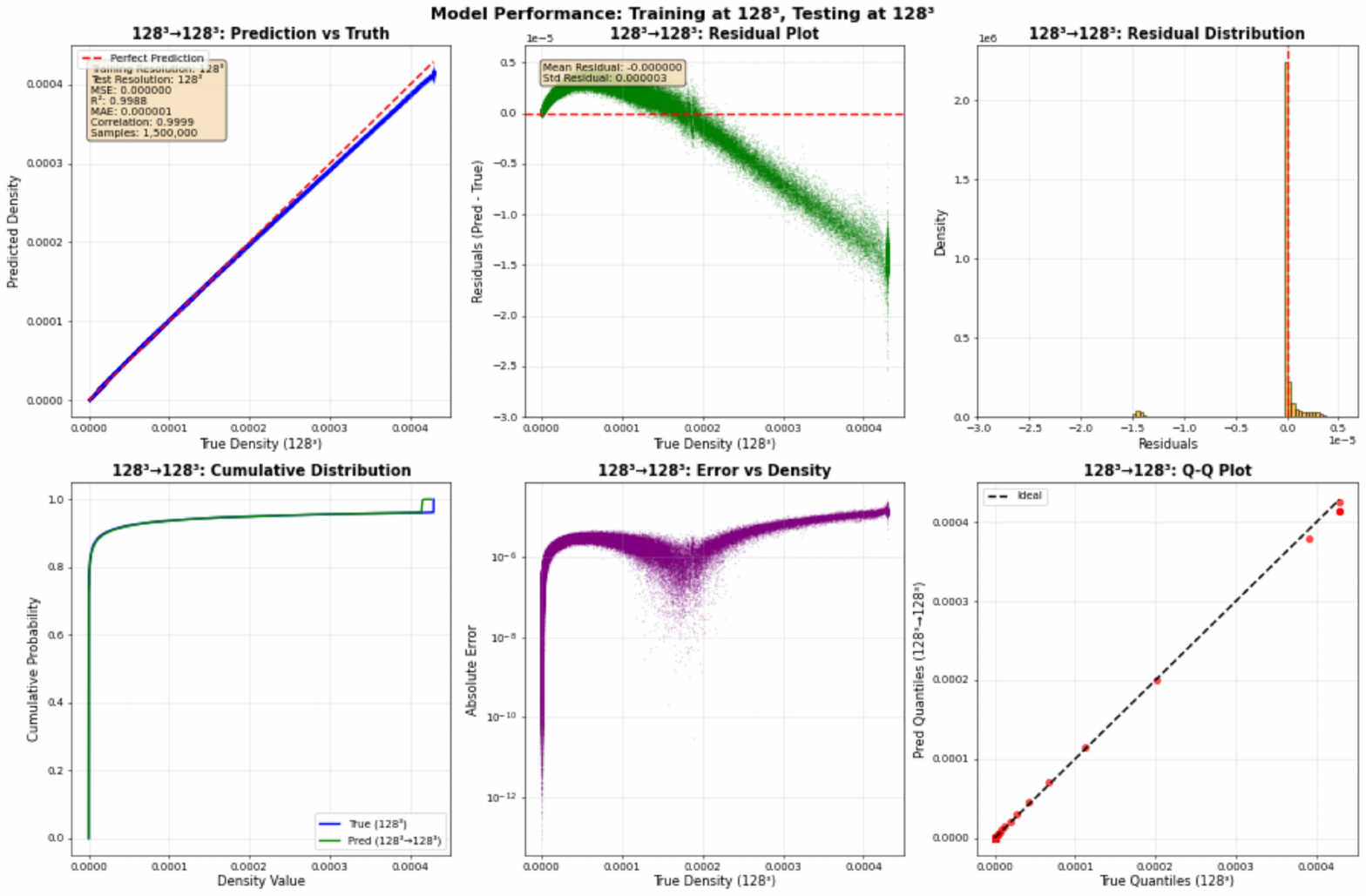}	
\caption{H2O correlation analysis.}	
\label{fig:resolution1}
\end{figure}

\paragraph{(b) Random molecular generalization within QM9.}
We next consider a substantially more demanding setting using the QM9 dataset.
A total of 100 molecules are randomly selected as a test set, with the remaining
molecules used for training. Unlike trajectory-based data, this split introduces
variations in molecular size, geometry, and bonding topology. To further assess
generalization beyond structural similarity, we analyze the distribution of local
bonding environments and identify test molecules containing bonding motifs that
do not appear anywhere in the training set.

%qm9 有三个分子和训练集分子局域成键环境不一样， 密度分布图直观一致， 相关性高， 误差低  ， 给出分子对比图， 找到训练集里的最相似结构
% 四个图， 0. 分子对比图   1. 直观密度对比图， 2. 误差图  3. 分辨率变化图 
\begin{figure}[!htb]   	
\centering	
\includegraphics[width=1.0\linewidth, scale=1]{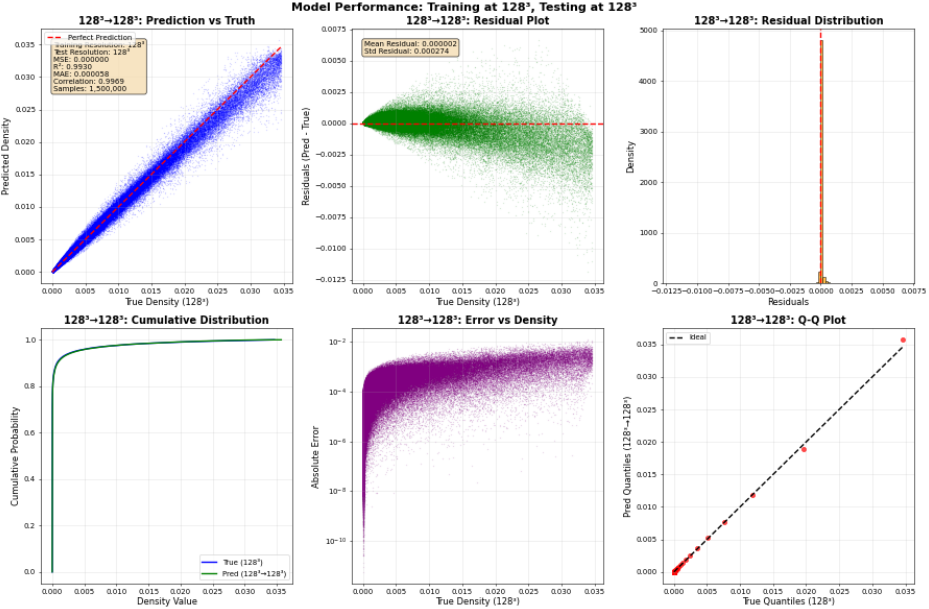}	
\caption{ Bonding environments extrapolation:  training on the first 5,000 molecules from QM9, and evaluation on molecules whose local bonding environments are not present in the training set, followed by correlation and error analysis between predicted densities and DFT-calculated densities.}	
\label{fig:resolution2}
\end{figure}

\begin{figure}[!htb]   	
\centering	
\includegraphics[width=1.0\linewidth, scale=1]{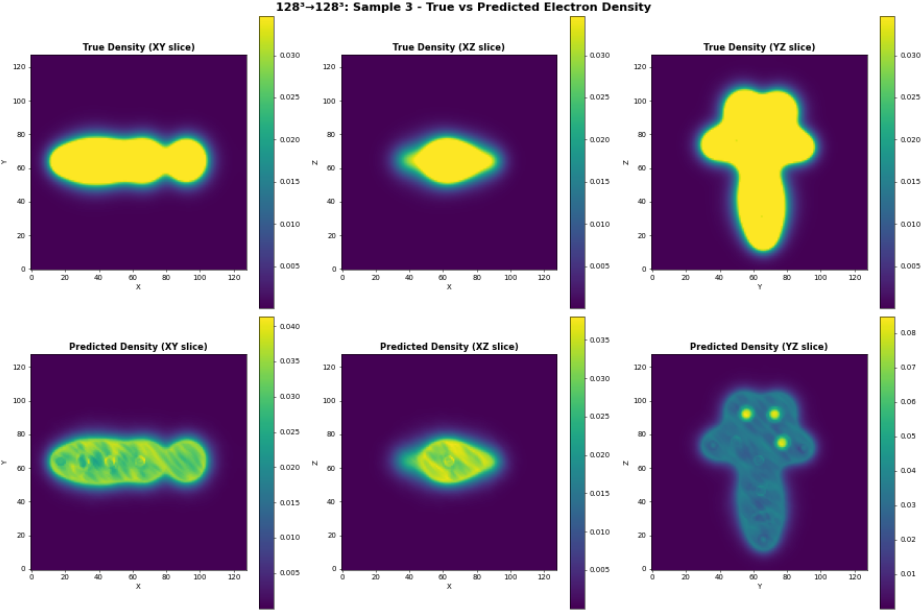}	
\caption{bonding enviroments generation:Comparison between predicted density and DFT-calculated density}	
\label{fig:resolution3}
\end{figure}

Despite the absence of these local environments during training, V2Rho-FNO
continues to predict physically meaningful electron densities for the test
molecules. Density isosurfaces and correlation analyses show strong agreement
with reference densities, although the overall test loss is higher than in the
trajectory-based case. This behavior reflects the broader sampling of the
external-potential function space and demonstrates that the model generalizes
beyond simple interpolation over known bonding patterns.

\paragraph{(c) Element-level extrapolation to fluorine-containing molecules.}
Finally, we assess a deliberately out-of-distribution scenario by training the
model exclusively on QM9 molecules containing only C, H, O, and N atoms, and
testing it on molecules that include fluorine. Under this split, all fluorine-related
bonding motifs and local chemical environments are absent from the training set.
Nevertheless, V2Rho-FNO produces density predictions that remain qualitatively
correct and exhibit substantial correlation with reference DFT densities. At the
same time, the test loss increases by approximately an order of magnitude
compared to the random QM9 split, indicating a controlled degradation in
accuracy under genuine chemical extrapolation.

%qm9  外推到 含F基团， 模型没有崩， 直观的 预测 密度 和DFT计算密度 直观观察具有非常高的一致性，  细致的相关性分析和误差分析， 它的相关性下降10%， 误差大一个数量级
% 两个图，  1. 直观密度对比图， 2. 误差图 ； 不需要分子对比图，只要文字阐述， 训练集里没有含氟分子， 测试集全部都是含氟分子即可

\begin{figure}[!htb]   	
\centering	
\includegraphics[width=1.0\linewidth, scale=1]{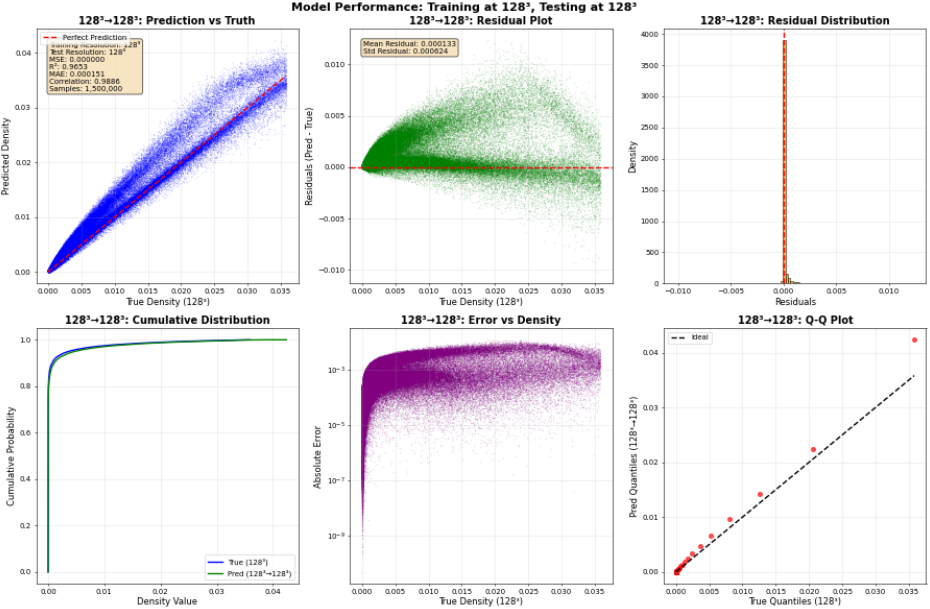}	
\caption{ Element-wise extrapolation of model-predicted densities: training on the remaining systems after excluding fluorine-containing groups from the first 5,000 molecules in QM9, followed by extrapolation to fluorine-containing molecules for prediction, and correlation and error analysis between predicted densities and DFT-calculated densities.}	
\label{fig:resolution4}
\end{figure}

\begin{figure}[!htb]   	
\centering	
\includegraphics[width=1.0\linewidth, scale=1]{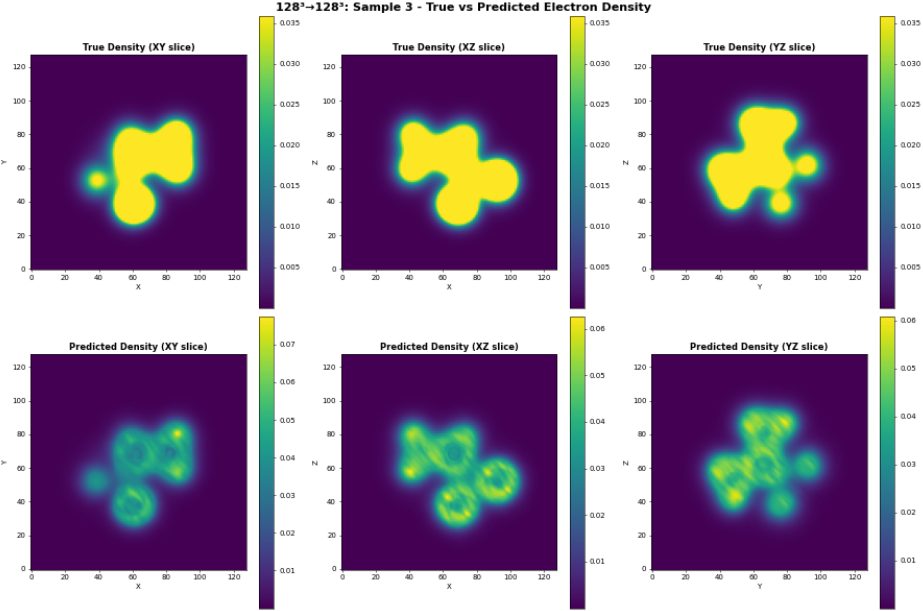}	
\caption{Element-wise extrapolation: Comparison between predicted density and DFT-calculated density}	
\label{fig:resolution5}
\end{figure}

Taken together, these three scenarios establish a clear hierarchy of
generalization difficulty: interpolation along a single trajectory yields the
highest accuracy, random molecular generalization across QM9 introduces moderate
error growth, and element-level extrapolation presents the most stringent test,
yet remains tractable within the operator-learning framework.

\subsection{Resolution transfer across training--test regimes}

A distinctive feature of the Fourier neural operator is its ability to evaluate
the learned mapping on spatial grids of different resolutions without retraining.
We assess this property by training models on coarse grids and performing inference
on finer grids via spectral zero-padding.

\begin{figure}[!htb]   	
\centering	
\includegraphics[width=1.0\linewidth, scale=1]{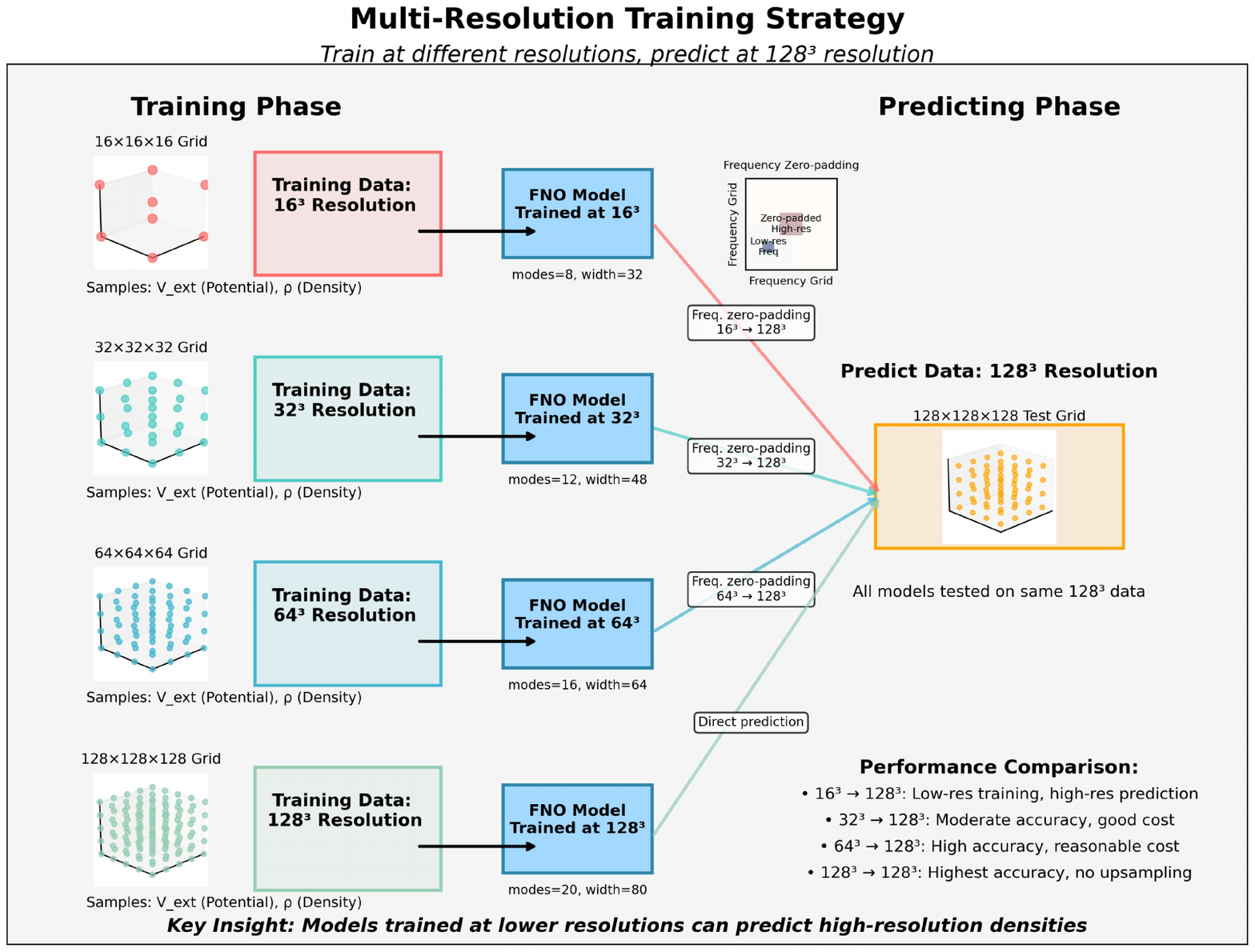}	
\caption{multi-resolution training scheme.}	
\label{fig:resolution6}
\end{figure}

\begin{figure}[!htb]   	
\centering	
\includegraphics[width=1.0\linewidth, scale=1]{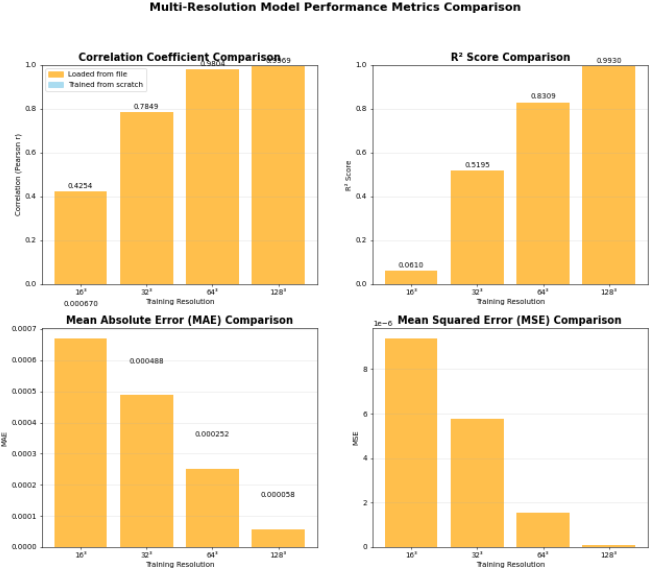}	
\caption{correlation and error analysis for different resolution training data on QM9.}	
\label{fig:resolution7}
\end{figure}

\begin{figure}[!htb]   	
\centering	
\includegraphics[width=1.0\linewidth, scale=1]{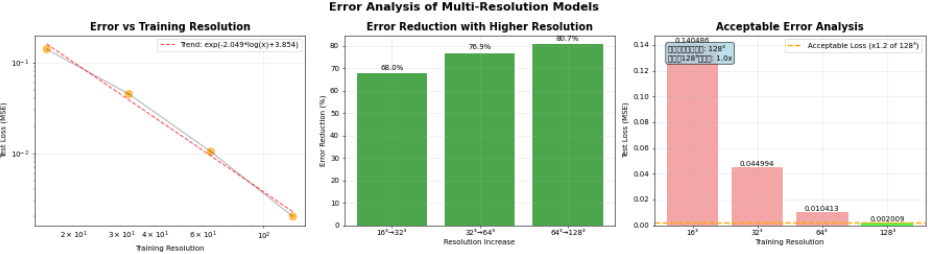}	
\caption{error ananysis for different resolution of training data on QM9.}	
\label{fig:resolution8}
\end{figure}
Across all three training--test relations described above, models trained at
moderate resolutions are able to generate physically consistent high-resolution
electron densities. In trajectory-based settings, even relatively coarse training
grids yield accurate super-resolved densities, reflecting the smooth nature of the
underlying potential variations. For the QM9 splits, models trained at 64$^3$
resolution produce high-fidelity predictions on 128$^3$ grids, while 32$^3$ models
capture the overall density distribution with partial smoothing of fine-scale
features. Models trained at 16$^3$ resolution consistently fail to reproduce
accurate high-resolution densities, indicating insufficient spectral bandwidth
to resolve chemically relevant length scales.

These observations confirm that resolution transfer in V2Rho-FNO follows a
band-limited behavior consistent with Fourier neural operator theory: once the
training resolution resolves the dominant spatial frequencies of the external
potential--density mapping, the learned operator can be evaluated on finer grids
without retraining.

\section{Discussion}

This work frames electron density prediction as an operator learning problem that targets an effective external-potential-to-density mapping consistent with pseudopotential Kohn--Sham DFT\cite{Kohn1965,Payne1992_RMP,Vanderbilt1990_USPP}. By using a screened real-space ionic potential as input and a three-dimensional Fourier neural operator (FNO) as the model backbone\cite{Li2021_FNO}, V2Rho-FNO establishes a direct connection between neural operator learning and the physical structure of density functional theory, while enabling discretization transfer via spectral evaluation.

%\paragraph{Operator interpretation via nonlocal density response.}
\subsection{Operator interpretation via nonlocal density response.}
From a physical standpoint, the response of the ground-state electron density to changes in the external potential is inherently nonlocal. In the linear-response regime, density variations obey
\begin{equation}
\delta \rho(\mathbf r)
=
\int \chi(\mathbf r,\mathbf r')\,\delta V_{\mathrm{ext}}(\mathbf r')\,\mathrm d\mathbf r'
+ \mathcal{O}(\delta V^2),
\label{eq:discussion_response}
\end{equation}
where $\chi(\mathbf r,\mathbf r')$ is the density response function \cite{Giuliani2005_ElectronLiquid}. Equation~\eqref{eq:discussion_response} already has the structure of an integral operator acting on the external potential. FNO layers implement precisely this class of nonlocal operators by combining global spectral convolutions with local nonlinear transformations \cite{Li2021_FNO,Kovachki2023_JMLRNeuralOperator}. Consequently, V2Rho-FNO can be interpreted as learning an effective, generally nonlinear generalization of the response kernel, providing a physically grounded view of how a neural operator can approximate the potential-to-density relation beyond linear response.

\subsection{Why a potential-field input is natural for neural operators.}
A key design choice is to represent each system by an ionic potential field rather than by discrete atom types and coordinates. Under the Hohenberg--Kohn theorem, the ground-state density is uniquely determined by the external potential (up to an additive constant) for fixed particle number \cite{Hohenberg1964}. In practical plane-wave pseudopotential calculations, the effective ionic potential entering the Kohn--Sham equations is smooth and numerically stable in real space\cite{Payne1992_RMP,Vanderbilt1990_USPP}, and provides a deterministic field representation of the system under fixed pseudopotential parameters and boundary conditions. Treating both input and output as scalar fields on the same spatial domain aligns directly with the function-to-function setting of neural operators, avoiding the need for manually designed embeddings or message-passing mechanisms to encode symmetries and long-range effects. This field-based formulation also provides a clear physical interpretation of the learned mapping and supports systematic analyses of discretization dependence.

\subsection{Origin of transferability across local bonding environments and chemical elements}

A central observation in the QM9 benchmarks is that V2Rho-FNO generalizes robustly to
molecular structures whose local bonding environments, and even constituent elements,
are entirely absent from the training set. Importantly, this behavior does not arise from
explicit encoding of chemical heuristics, but follows naturally from the operator-based
formulation of the model.

From the perspective of operator learning, the training data are not treated as discrete
samples indexed by molecular identity, but as realizations of an underlying functional
relationship
\begin{equation}
\mathcal{G}:\; V_{\mathrm{ext}}(\mathbf r)\;\mapsto\;\rho(\mathbf r),
\label{eq:operator_map_discussion}
\end{equation}
defined on a space of physically admissible external potentials. In this setting, different
local bonding environments and chemical compositions correspond to different regions of
the input function space, rather than to qualitatively distinct categories. Generalization
is therefore governed by how well the training set spans this function space, rather than
by structural similarity between individual molecules.

The Fourier neural operator plays a crucial role in enabling this behavior. Each FNO layer
implements a global convolution in Fourier space,
\begin{equation}
(\mathcal{K}v)(\mathbf r)
=
\mathcal{F}^{-1}\!\left(
R(\mathbf k)\,\mathcal{F}[v](\mathbf k)
\right),
\label{eq:fno_kernel}
\end{equation}
where $R(\mathbf k)$ denotes a learned spectral multiplier and $\mathcal{F}$ the Fourier
transform. This construction defines a translation-equivariant, nonlocal operator whose
action is determined by the spectral content of the input field, rather than by discrete
atomic labels or predefined bonding patterns. As a result, the same learned operator acts
uniformly on all input potentials within its domain of approximation.

Crucially, changes in chemical composition or local bonding motifs manifest as structured
variations in the external potential field, including shifts in amplitude, range, and
spatial gradients. Provided that the training data encompass a sufficiently rich set of
such variations, the learned operator can interpolate and extrapolate across local bonding
environments without requiring explicit exposure to every possible chemical motif. This
explains why V2Rho-FNO can accurately predict electron densities for molecules containing
previously unseen local bonding environments or chemical elements, as observed in the
QM9 splitting schemes.

From this viewpoint, the observed transferability is not an empirical artifact, but a
direct consequence of learning a function-to-function operator that is independent of
molecular identity. In the limit of sufficient model capacity and adequate coverage of
the input potential space, the learned operator increasingly approximates a universal
external-potential-to-density map, consistent with the universality implied by the
Hohenberg--Kohn theorem.

Interpreting density prediction through response theory provides a physically grounded explanation of why a Fourier neural operator (FNO) is a natural architecture for learning the HK map.
In exact DFT, the density--potential mapping has a rich functional-analytic structure, and whenever the HK map $\mathcal{G}$ is differentiable at a potential $V_0$, its derivative is the static density response operator $\chi_{V_0}$ \cite{Penz2022_StructureHK,Giuliani2005_ElectronLiquid}.
That is, the HK map admits the local expansion
\begin{equation}
\mathcal{G}[V_0+\delta V]
=
\mathcal{G}[V_0]
+
\int \chi_{V_0}(\mathbf r,\mathbf r')\,\delta V(\mathbf r')\,d\mathbf r'
+
\mathcal{O}(\delta V^2),
\end{equation}
revealing that the \emph{native local geometry} of the HK map is that of a nonlocal integral operator.
FNO layers implement precisely such nonlocal operators by learning spectral multipliers that parameterize convolution-type kernels, combined with pointwise nonlinearities \cite{Li2021_FNO,Kovachki2023_JMLRNeuralOperator}.
Importantly, this view is fully consistent with (and clarifies) ``system-independent'' generalization: one does not learn a single system-specific $\chi$, but rather a nonlinear operator that reproduces response-like behavior across the space of admissible external potentials, with generalization controlled by function-space coverage rather than sample-level similarity \cite{Kovachki2023_JMLRNeuralOperator}.

\subsection{Resolution transfer and spectral zero-padding in Fourier neural operators}
A distinctive advantage of the Fourier neural operator formulation is that it learns a truncated Fourier representation of an underlying continuous operator. As a result, the learned mapping is inherently band-limited: it accurately represents the system response only within the spectral bandwidth resolved during training. Within this framework, resolution transfer from a coarse grid to a finer grid can be achieved by spectral zero-padding. After evaluating the learned operator on the available Fourier modes, the spectral representation is embedded into a higher-resolution Fourier grid by padding the unresolved high-frequency components with zeros, followed by an inverse Fourier transform to real space. Mathematically, this corresponds to a band-limited interpolation of the learned continuous operator rather than an extrapolation beyond its spectral support.

From the perspective of electronic structure, this behavior is physically transparent. Low and intermediate spatial frequencies in the density encode smooth charge redistribution and long-range electrostatic response, which are well captured on coarse grids. In contrast, sharp near-core variations correspond to intrinsically high-frequency physics and are either suppressed by the screened potential representation or intentionally excluded from the learning objective when near-core density values are truncated. It is therefore expected that super-resolved densities generated by spectral zero-padding reproduce the correct large-scale structure and maintain strong agreement with reference densities at low and intermediate frequencies, while deviations may remain in regions dominated by high-frequency features. Importantly, this resolution transfer should be interpreted as a controlled, physically consistent band-limited extension, rather than a reconstruction of missing high-frequency information.

\subsection{Relation to recent neural-operator-based density models}
Recent work has begun to explore neural-operator concepts for density estimation from atomistic inputs. A closely related example is the Gaussian Plane-Wave Neural Operator (GPWNO), which predicts electron densities from discrete atomic structures by combining operator-inspired updates with hybrid plane-wave and Gaussian representations \cite{Kim2024_GPWNO}. GPWNO formulates density estimation as a mapping from element types and coordinates to continuous densities, employing probe points and basis-level inductive biases to capture multiscale features. In contrast, V2Rho-FNO adopts a field-to-field formulation that directly targets an effective external-potential-to-density mapping: the input is an ionic potential field, and the output is the electron density on a uniform grid. This avoids explicit atom embeddings and yields a representation that is naturally compatible with the neural operator paradigm.

Methodologically, V2Rho-FNO emphasizes discretization transfer as a first-class capability enabled by the spectral structure of FNOs, allowing inference on finer grids via spectral zero-padding without retraining. While GPWNO introduces strong physics-motivated inductive biases that may benefit localized high-frequency features, our approach provides a simpler operator-level formulation grounded in potential-field inputs and is particularly suited for grid-based workflows where resolution transfer and operator interpretation are central.

\subsection{Outlook}

The operator-learning perspective developed here suggests several natural extensions. 
First, while the present study focuses on molecular systems in finite simulation boxes, the same formulation can be generalized to periodic crystalline systems by adopting periodic boundary conditions and lattice-consistent representations of the ionic potential. This would enable neural-operator-based density prediction for extended solids within a fully first-principles framework.

Second, the V2Rho-FNO framework can be naturally extended to the prediction of \emph{spin density}, thereby enabling applications to magnetic and 
spin-polarized systems. Within spin-density functional theory, the external potential---augmented by spin-dependent contributions---determines both the spin-up and spin-down 
electron densities. From an operator-learning standpoint, this corresponds to a straightforward generalization from a scalar-valued density field to a vector-valued spin density field. Such an extension would allow the present approach to address open-shell molecules, magnetic materials, and systems with nontrivial spin polarization, providing direct access to spatially resolved magnetic information without altering the underlying operator architecture.

Finally, the operator viewpoint is not restricted to ground-state densities. By incorporating quasiparticle corrections, the same framework could be coupled with many-body perturbation theory, such as the $GW$ approximation, to learn effective mappings from ionic potentials to excited-state or quasiparticle densities. This opens 
the possibility of data-driven acceleration of excited-state electronic structure calculations within a unified operator-learning paradigm.

\section{Methods}

\paragraph{HK map as a response-type operator and the FNO hypothesis class.}
We consider an operator-learning formulation of the (effective) HK map,
\begin{equation}
\mathcal{G}:\; V_{\mathrm{ext}}(\mathbf r)\in\mathcal{V}\;\longmapsto\;\rho(\mathbf r)\in\mathcal{R},
\end{equation}
where $\mathcal{V}$ and $\mathcal{R}$ denote suitable function spaces on the simulation domain.
The density--potential mapping has been analyzed in detail in the mathematical foundations of DFT, including its regularity and differentiability properties \cite{Penz2022_StructureHK}.
Where $\mathcal{G}$ admits a local derivative at $V_0$, the first-order variation is given by the static density response operator,
\begin{equation}
D\mathcal{G}[V_0](\delta V)(\mathbf r)
=
\int \chi_{V_0}(\mathbf r,\mathbf r')\,\delta V(\mathbf r')\,d\mathbf r',
\label{eq:method_frechet_chi}
\end{equation}
which is the standard linear-response relation in many-body theory and (time-dependent) DFT \cite{Giuliani2005_ElectronLiquid,Ullrich_TDDFT,MarquesGross2004_TDDFT}.

\paragraph{Remark on differentiability and response structure.}
We note that, in exact density functional theory, the density--potential mapping may fail
to be differentiable on parts of its domain \cite{Penz2022_StructureHK}. Nevertheless,
it has been shown that differentiable-but-exact regularized formulations exist, in which
the Hohenberg--Kohn map becomes Fr\'echet differentiable while remaining physically
equivalent in the appropriate limit \cite{Kvaal2013_DifferentiableDFT}. Within such a
framework, the linear-response operator $\chi$ can be rigorously interpreted as the local
derivative of the HK map, providing a well-defined response-theoretic motivation for the
neural-operator hypothesis class adopted here.

Since exact DFT functionals may be nondifferentiable on parts of their domain, we also note that differentiable-but-exact regularized formulations exist (e.g., via Moreau--Yosida regularization), in which the relevant functionals become Fr\'echet differentiable while remaining exact in the appropriate limit \cite{Kvaal2013_DifferentiableDFT}.
Motivated by the intrinsic nonlocal operator structure in Eq.~\eqref{eq:method_frechet_chi}, we adopt Fourier neural operators as a hypothesis class, which are universal approximators for nonlinear operators between function spaces and are discretization-invariant by construction \cite{Kovachki2023_JMLRNeuralOperator,Li2021_FNO}.

\subsection{Problem formulation and physical setting}

All calculations are performed within the Born--Oppenheimer approximation \cite{BornOppenheimer1927}, where the ionic configuration is fixed and the electronic ground state is uniquely determined by the external electrostatic potential $V_{\mathrm{ext}}(\mathbf{r})$.
For a given atomic structure, the external potential is constructed as the superposition of nuclear Coulomb potentials and serves as the sole input to the machine-learning model. The target output is the corresponding ground-state electron density $\rho(\mathbf{r})$ obtained from Kohn--Sham density functional theory (DFT) calculations \cite{Kohn1965}, typically in a plane-wave pseudopotential framework \cite{Payne1992_RMP,Vanderbilt1990_USPP}.

The learning task is formulated as a field-to-field regression problem, in which a neural operator approximates the mapping
\begin{equation}
\mathcal{G}: V_{\mathrm{ext}}(\mathbf{r}) \mapsto \rho(\mathbf{r}),
\end{equation}

defined on a three-dimensional spatial domain discretized on uniform real-space grids. Rather than learning pointwise correlations on a fixed discretization, the objective is to approximate a resolution-invariant operator that generalizes across spatial resolutions.

\subsection{Data representation and preprocessing}

The external potential and electron density are sampled on cubic grids of size $N \times N \times N$, with $N \in \{16, 32, 64, 128\}$, depending on the training resolution. All systems are embedded in a fixed simulation box, and periodic boundary conditions are not assumed unless otherwise stated.

To ensure numerical stability during training, the electron density in the immediate vicinity of nuclear singularities is truncated to a fixed upper bound. This truncation is applied consistently to both reference densities and model predictions and is used exclusively to stabilize optimization on uniform grids. Regions affected by truncation are excluded from all quantitative error metrics and correlation analyses, ensuring that reported accuracies reflect physically meaningful density variations away from the nuclear cores.

Both input potentials and output densities are normalized using statistics computed from the training set.

\subsection{Fourier neural operator architecture}

To approximate the external-potential--to--density mapping, we employ a three-dimensional Fourier neural operator (FNO). The FNO represents operators in a discretization-invariant manner by combining global spectral representations with local real-space processing.

\begin{figure}[!htb]   	
\centering	
\includegraphics[width=1.0\linewidth, scale=1]{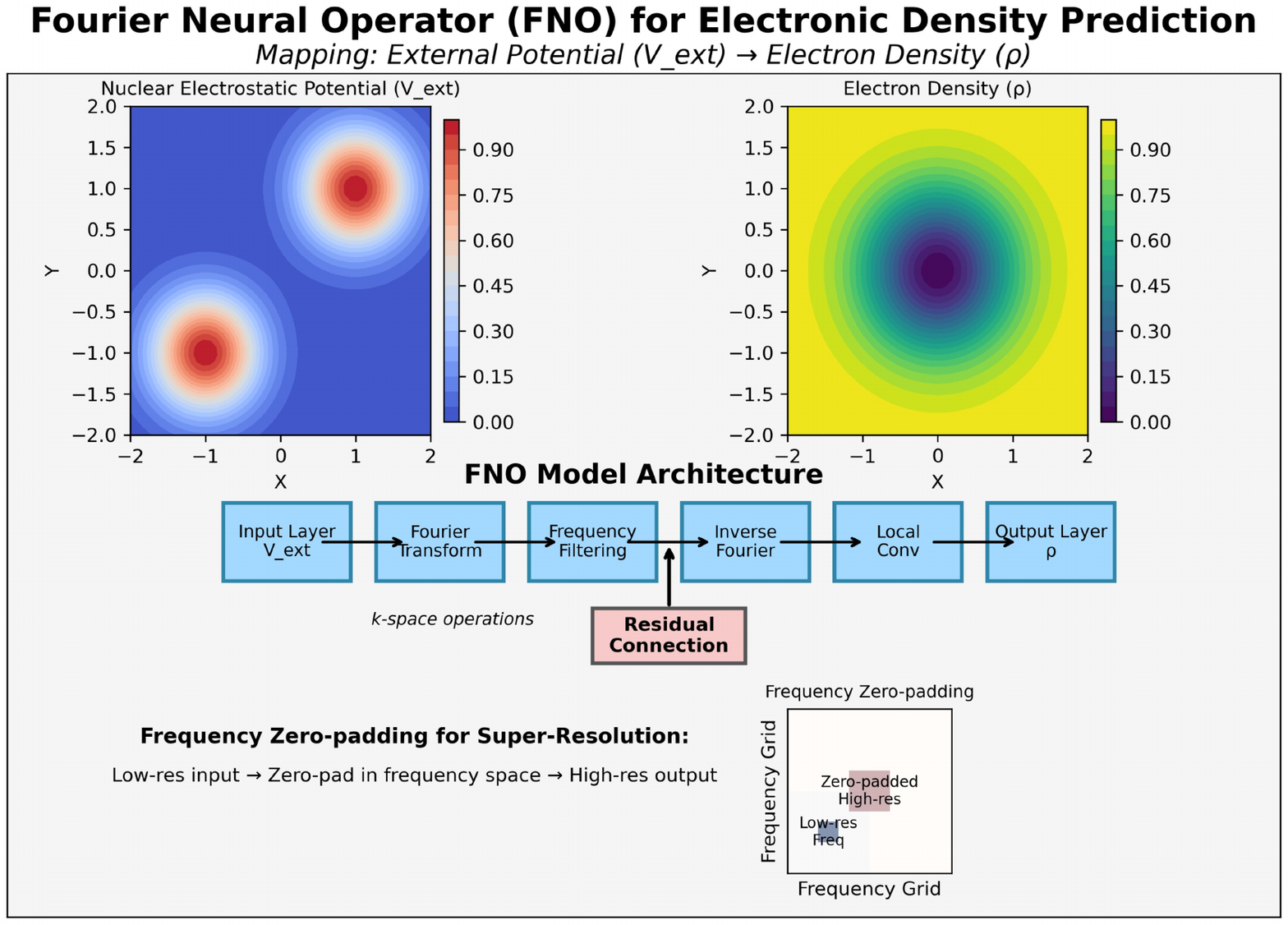}	
\caption{multi-resolution training scheme.}	
\label{fig:resolution9}
\end{figure}

\subsubsection{Input lifting}

The scalar external potential field $V_{\mathrm{ext}}(\mathbf{r})$ is first mapped to a higher-dimensional latent representation through a linear embedding followed by a nonlinear activation,
\begin{equation}
u_0(\mathbf{r}) = \sigma \left( W_{\mathrm{in}} V_{\mathrm{ext}}(\mathbf{r}) + b_{\mathrm{in}} \right),
\end{equation}
where $W_{\mathrm{in}}$ projects the input to a latent channel dimension referred to as the model width.

\subsubsection{Fourier layers}

The core of the network consists of multiple stacked Fourier layers. Each Fourier layer performs the following operations:

\begin{enumerate}
    \item \textit{Spectral transform.}  
    The latent field is transformed from real space to Fourier space using a real-to-complex fast Fourier transform (FFT).

    \item \textit{Low-frequency spectral mixing.}  
    Only a fixed number of low-frequency Fourier modes are retained along each spatial dimension. These modes are linearly transformed using learned complex-valued weights, while all higher-frequency components are discarded. This restriction enforces a band-limited representation and focuses the model capacity on long-range, nonlocal interactions.

    \item \textit{Inverse transform.}  
    The modified spectral coefficients are transformed back to real space via an inverse FFT.

    \item \textit{Local real-space update.}  
    A pointwise or small-kernel convolution is applied in real space to capture short-range, localized features that are not efficiently represented in the spectral domain.

    \item \textit{Residual connection and nonlinearity.}  
    The output of each Fourier layer is combined with its input through a residual connection, followed by normalization and a nonlinear activation function.
\end{enumerate}

This hybrid spectral--local design enables the FNO to simultaneously capture global electronic responses to the external potential and localized density variations near atomic environments.

\subsubsection{Output projection}

After the final Fourier layer, the latent representation is projected back to a scalar field representing the electron density through a multilayer perceptron applied pointwise in real space. A strictly positive activation function is used in the final layer to enforce the non-negativity of the predicted density.

\subsection{Multi-resolution training strategy}

To investigate the resolution-invariance of the learned operator and to reduce computational cost, FNO models with identical architecture are trained independently on datasets discretized at different spatial resolutions ($16^3$, $32^3$, $64^3$, and $128^3$).

Each model is trained exclusively on data at its native resolution and does not access higher-resolution information during training. Model parameters, including the number of retained Fourier modes and channel width, are scaled moderately with resolution to ensure stable optimization.

\subsection{Resolution lifting via frequency-domain zero-padding}

During inference, models trained at lower resolutions are evaluated on a higher-resolution grid by exploiting the resolution-invariant formulation of the Fourier neural operator. Specifically, after the spectral transform, the learned low-frequency Fourier coefficients are embedded into a larger Fourier grid corresponding to the target resolution by zero-padding all higher-frequency components. An inverse Fourier transform then yields predictions on the finer real-space grid.

This frequency-domain zero-padding does not introduce new high-frequency information beyond what is learned during training. Instead, it produces a smooth spectral extension of the band-limited representation, analogous to standard Fourier interpolation. As a result, resolution lifting yields high-resolution density predictions that are free from spurious fine-scale artifacts.

\subsection{Training procedure}

All models are trained using a mean-squared-error loss computed on the electron density field, excluding grid points affected by density truncation near nuclear positions. Optimization is performed using the Adam optimizer with a fixed learning rate schedule. Training is terminated once convergence on a held-out validation set is achieved.

\subsection{Evaluation protocol}

To enable a controlled comparison across resolutions, all models---regardless of training resolution---are evaluated on the same high-resolution ($128^3$) test set. For models trained at lower resolutions, predictions are first lifted to the target resolution using frequency-domain zero-padding. Error metrics and correlation coefficients are computed consistently on the lifted predictions and reference densities, excluding truncated regions.

This evaluation protocol allows for a systematic assessment of the trade-off between training resolution, computational efficiency, and prediction accuracy.

\subsection{Implementation details}

All models are implemented in PyTorch. Three-dimensional FFT operations are performed using the built-in real-to-complex FFT routines. Training and inference are carried out on NVIDIA GPUs. Hyperparameters, including the number of Fourier layers, retained modes, and channel widths, are reported in the Supplementary Information.

\subsection{Quantifying function-space coverage and generalization}

To analyze the generalization behavior of V2Rho-FNO from an operator-learning
perspective, we introduce a set of metrics that quantify (i) coverage of the
external-potential function space by the training data and (ii) the resulting
prediction error on unseen inputs. These metrics are explicitly aligned with
the spectral structure of the Fourier neural operator.

\subsubsection{FNO-aligned embedding of the external potential}

The Fourier neural operator represents input functions through a truncated
spectral expansion, retaining only a finite number of low-frequency Fourier
modes \cite{Li2021_FNO,Kovachki2023_JMLRNeuralOperator}. Accordingly, we define
an embedding of the external potential that reflects the degrees of freedom
resolved by the model.

Given an external potential $V(\mathbf r)$ defined on a uniform spatial grid,
we compute its Fourier transform
\begin{equation}
\hat V(\mathbf k) = \mathcal{F}[V](\mathbf k),
\end{equation}
and construct a low-frequency embedding
\begin{equation}
\Phi(V) =
\left\{
\left| \hat V(\mathbf k) \right|
\;\middle|\;
\|\mathbf k\| \le k_{\mathrm{cut}}
\right\},
\label{eq:phi_embedding}
\end{equation}
where $k_{\mathrm{cut}}$ denotes the spectral cutoff determined by the model
resolution and the number of retained Fourier modes. This embedding extracts
the magnitude of the low-frequency spectral components of the external potential
and provides a finite-dimensional representation of the input function aligned
with the FNO hypothesis class.

\subsubsection{Function-space coverage distance}

To quantify how well the training data cover the relevant input function space,
we define a coverage distance for each test potential $V^{(t)}$ as its distance
to the nearest training potential in the embedding space:
\begin{equation}
d_{\mathrm{cov}}(V^{(t)}) =
\min_{V^{(i)} \in \mathcal{D}_{\mathrm{train}}}
\left\|
\Phi\!\left(V^{(t)}\right)
-
\Phi\!\left(V^{(i)}\right)
\right\|_2 .
\label{eq:coverage_distance}
\end{equation}
This quantity measures how far a test input lies from the support of the training
set within the low-frequency subspace resolved by the neural operator. Smaller
values of $d_{\mathrm{cov}}$ indicate strong coverage and interpolation-like
generalization, whereas larger values correspond to extrapolation in the input
function space.

\subsubsection{Prediction error metrics}

Model accuracy is evaluated using masked error metrics that exclude near-core
grid points where the reference density is truncated by construction. For each
test sample, we compute the mean squared error (MSE),
\begin{equation}
\mathrm{MSE} =
\frac{1}{|\Omega|}
\sum_{\mathbf r \in \Omega}
\left(
\rho_{\mathrm{pred}}(\mathbf r)
-
\rho_{\mathrm{ref}}(\mathbf r)
\right)^2 ,
\label{eq:mse}
\end{equation}
and the Pearson correlation coefficient between the predicted and reference
densities over the same spatial domain $\Omega$.

\subsubsection{Coverage--error analysis}

By analyzing the joint distribution of the coverage distance
$d_{\mathrm{cov}}(V^{(t)})$ and the corresponding prediction error, we assess
how generalization performance depends on coverage of the external-potential
function space. This analysis enables a systematic comparison of different
training--test relations, ranging from interpolation within narrowly sampled
subspaces to element-level extrapolation involving previously unseen potential
features, and provides an empirical link between operator-level approximation
theory and observed prediction accuracy.

\subsection{Summary}

Overall, the proposed framework treats electron density prediction as a resolution-invariant operator learning problem, enabling efficient generalization across spatial scales while preserving physically meaningful nonlocal responses to the external potential.

%\section{Data Availability}
%All molecular structures and corresponding electron density data generated in this study are archived in [Database Name] and publicly available at [URL].
%\section*{Code Availability}
%The PyTorch implementation of FNO for electronic density prediction and associated analysis tools are available at [GitHub Repository URL] under the MIT license.

\section{Acknowledgments}
This work was supported by Innovation Program for Quantum Science and Technology (2021ZD0303306), the National Natural Science Foundation of China (22393913, 22422304, 22288201), the Strategic Priority Research Program of the Chinese Academy of Sciences (XDB0450101).

% ====== bibliography hook (BibTeX) ======
%\bibliographystyle{unsrt}
\bibliography{references}

\end{document}